\documentclass{article}

\usepackage{arxiv}

\usepackage[utf8]{inputenc} 
\usepackage[T1]{fontenc}    
\usepackage{hyperref}       
\usepackage{url}            
\usepackage{booktabs}       
\usepackage{amsfonts}       
\usepackage{nicefrac}       
\usepackage{microtype}      
\usepackage{lipsum}		
\usepackage{graphicx}
\usepackage{doi}
\usepackage{amsmath}
\usepackage{gensymb} 
\usepackage{subfig}

\title{The sign of risk for present value of future losses. }

\author{ \href{https://orcid.org/0000-0003-1202-2795}{\includegraphics[scale=0.06]{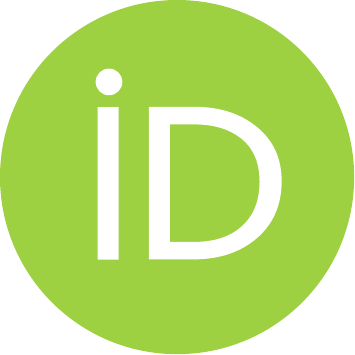}\hspace{1mm}Brian P. Hanley} \\
	Butterfly Sciences\\
	Post Falls, ID 83854 \\
	\texttt{brian.hanley@bf-sci.com} \\
	\And
	\href{}{\includegraphics[scale=0.06]{orcid.pdf}\hspace{1mm}Steve Keen} \\
	University College London \\ Institute for Security \& Resilience Studies\\
	\texttt{debunking@gmail.com} \\
}

\renewcommand{\headeright}{Preprint}
\renewcommand{\undertitle}{Preprint}

\hypersetup{
pdftitle={A template for the arxiv style},
pdfsubject={q-bio.NC, q-bio.QM},
pdfauthor={David S.~Hippocampus, Elias D.~Striatum},
pdfkeywords={First keyword, Second keyword, More},
}

\begin{document}
\maketitle

\begin{abstract}
In the ongoing debate over discount rates and climate change, William Nordhaus has championed a higher discount rate to account for risk. Nicholas Stern has championed a lower rate. Here we prove that in the case of a stream of future losses, risk can only be represented by a lower discount rate, never a higher one.
\end{abstract}


\section{Introduction}
Let us consider the meaning of present value ($PV$) for a future stream of losses ($PV_{FL}$), as opposed to income ($PV$). This discussion comes out of the long and vigorous discount debate in the context of climate economics. This debate can be exemplified by the discussion between William Nordhaus \cite{Nordhaus2007CriticalAssumptionsinSternReviewClimate}, and Nicholas Stern \cite{Stern2007EconomicsofClimateChangeSternReview}\cite{Howard2017Meta-analysisOFClimateDamageEstimates}. Here we will show that both sides of this debate have missed a key matter.  

For our presentation here, we accept the nominal use of discounting a future loss stream, but note the ethics debate on inter-generational equity that questions the morality of discounting climate losses at all \cite[p 24]{Stern2021TheEconomicsofImmenseRiskUrgentActionandRadicalChange}. 

Discounting of a future loss stream, like discounting of a future income stream is based on two aspects of discounting, the time value of money ($\mathbf{d_{TVM}}$) and the risk discount factor ($\mathbf{r_{rar}}$) that is applied to $\mathbf{d_{TVM}}$

To examine this matter of discounting $\mathbf{PV_{FL}}$ for risk, we will start with income, which is the usual case that $\mathbf{PV}$ was created for, step through that logic, then apply the same logic to a stream of future losses. We will begin with an expositional discussion, and then proceed to the mathematical proof. 

\section{Present value of future income}
When an evaluation is made of risk relative to a future stream of income, what is the risk an investor worries about? The risk is that real future income will be \emph{lower} than expected based on current discount rates. So, to address this, a projection is made using a \emph{higher} discount rate to show what the effect on PV is. The higher the discount rate, the \emph{lower} the PV becomes. 

\begin{figure}[!ht]
    \centering
    \subfloat[Present value of future income ($PV$) ] {{\includegraphics[width=5cm]{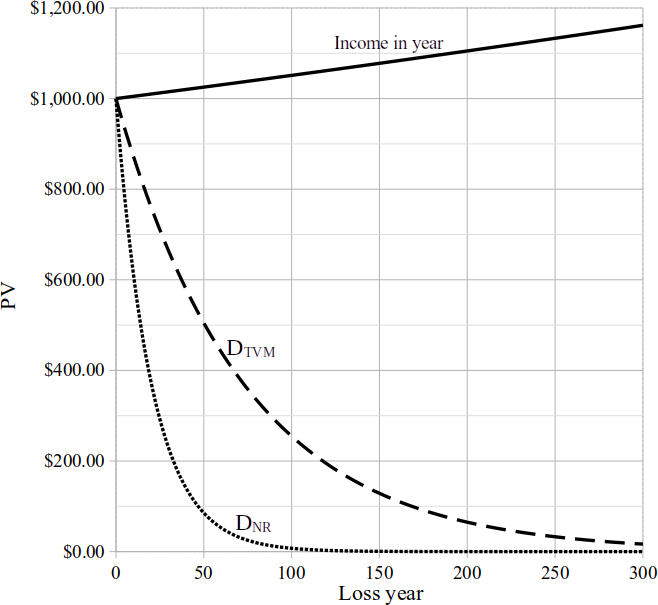} }}
    \quad
    \subfloat[Present value of future losses ($PV_{FL}$)] {{\includegraphics[width=5cm]{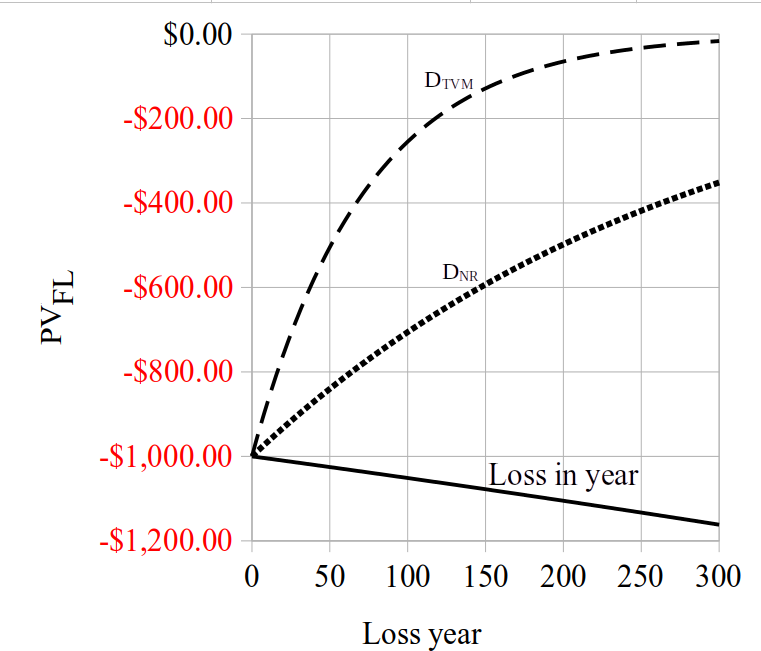} }}
    \caption{\textbf{Effects of discounting} Discounting of income (a) lowers the income because of cost of money and risk, with risk decreasing PV of income. Discounting losses (b) lowers the losses due to cost of money and risk, with risk increasing the losses relative to time value of money. Two discount rates are shown: $\mathbf{d_{TVM}}$ nominally set = 1.427\% and discounts modified by $\mathbf{r_{rar}}$.   
The term of 300 years was chosen because of DICE models terminating in the year 2300 \cite{dietz2021tipping}.  }
    \label{Fig_PV}
\end{figure}

For the purposes of this discussion, we will nominally assign the mean inflation adjusted federal reserve discount rate ($1.43\%$) to $\mathbf{d_{TVM}}$. The precise value of $\mathbf{d_{TVM}}$ can be varied. The PV curve for $\mathbf{d_{TVM}}$ is visible in fig. \ref{Fig_PV}(a).

To adjust a future projection of income for risk, this baseline rate, $\mathbf{d_{TVM}}$, will be modified by the risk discount factor, $\mathbf{r_{rar}}$, giving us a discount net of risk: $\mathbf{d_{NR}} = \mathbf{d_{TVM}} \times \mathbf{r_{rar}}$ eq. \ref{eq:PVRisk}. 

William Nordhaus used up to 5.1\% risk accounted discount rate \cite{Nordhaus2017RevisitingSCC_Supplement}. Using a 5.1\% DICE discount rate, $r_{_{DF}}={\frac{d_{DICE}}{d_{TVM}}} = 3.57$. This 3.57 risk adjustment factor we apply to $d_{TVM}$ to obtain a risk curve, for $d_{Risk}$. in fig. \ref{Fig_PV}(a).

\begin{align}
\label{eq:PVRisk}
    \begin{split} \text{(a) Calculation} & \text{ of } r_{rar} \text{ using DICE rate.} \\
r_{_{DF}}=& {\frac{d_{DICE}}{d_{TVM}}} = 3.57 \\ 
\text{Where: } d_{TVM}\, =& \text{ time value of money} \\ 
d_{DICE} =& \text{ DICE 2016R discount rate, 5.1\%} \\
r_{_{DF}} =& \text{ risk discount factor} 
    \end{split}
    \begin{split} 
    \text{(b) } PV \text{ discount } & \text{net of risk}\\
d_{NR} =& \; d_{TVM} \times r_{rar} \\
\text{Where: } d_{NR}\, =&\text{ discount net of risk } \\  \\ \\
    \end{split}    
\end{align}

\section{Present value of future losses ($\mathbf{PV_{FL}}$)} 
Now, let us ask the question, "What does risk mean for a stream of future losses?" Risk in this case means that the losses, in real terms, could be \emph{larger} than expected, rather than smaller. Note that this is the opposite of the meaning of risk for future income, although the direction of net income after accounting for risk is the same, as both curves move in a negative direction. So how can we represent losses being greater than expected using a discount rate? 

To represent larger losses using a discount rate, the rate must be \emph{lower} than than the baseline $\mathbf{d_{TVM}}$ rate. So, instead of \emph{multiplying} to increase the discount rate by  $\mathbf{r_{rar}}$ as is correct for PV of future income, one must divide by $\mathbf{r_{rar}}$. (Alternatively, a risk factor could be subtracted from the discount rate.) The curve for the $\mathbf{PV_{FL}}$ version of $\mathbf{d_{NR}}$ is shown in eq. \ref{eq:PVFLRisk} and fig. \ref{Fig_PV}(b). 

\begin{align}
    \begin{split} \label{eq:PVFLRisk}
    PV_{FL} & \text{ discount net of risk}\\
d_{NR} &= \frac{d_{TVM}} {r_{rar}} \\
\text{Where: } d_{NR}\, &=\text{ discount net of risk } \\        
    \end{split}
\end{align}

\section{A proof that risk adjustment for losses must have the opposite sign to the discount rate for the time value of money}
Let us begin with a project that has an expected positive revenue stream $E(t)$, and an actual positive revenue stream $A(t)$. At an arbitrary value of $t$, $E(t)$ is compared with $A(t)$. We will assume that both $E(t)$ and $A(t)$ start at the same initial value of income $I$, and grow exponentially at different rates $g_{_E}$ and $g_{_A}$ respectively. The actual and expected returns are:
\begin{align}
E(t)&=I \cdot e^{(g_{_E}) \cdot t}   &  A(t)&=I \cdot e^{(g_{_A}) \cdot t} 
 &\text{Where: } g_{_E} \text{ and } g_{_A} > 0
\end{align}

The discounted value of those cash flows subtracts a \emph{positive} term for the time value of money, $d_{tvm}$, from both exponents, to account for the lower present value of future income versus income today. 
\begin{align} \label{eq:Etvm_Atvm1}
E_{tvm}(t)&=I \cdot e^{(g_{_E} - d_{tvm} ) \cdot t}   &  A_{tvm}(t)&=I \cdot e^{(g_{_A} - d_{tvm}) \cdot t} 
& \text{Where: } E_{tvm}(t) > A_{tvm}(t) > 0 
\end{align}

Let us next consider the relevant risk in the ordinary present value case, that actual returns will be below expected returns. It is possible for a project with a positive future income stream to be net present value positive on a time value of money basis, but lower---or even negative---on a risk adjusted basis. Thus, a risk adjustment rate, $d_{rar}$, is applied to $E_{tvm}(t)$ to cover the case where returns are \emph{lower} than expected. There is no need to apply $d_{rar}$ to actual returns, because \emph{actual returns} are the basis for determining what the correct fit value of $d_{rar}$ is.
\begin{align} \label{eq:Etvm_Atvm2}
E_{rar}(t)&=I \cdot e^{(g_{_E} - d_{tvm} - d_{rar} ) \cdot t}   &  A_{rar}(t) &= A_{tvm}(t) =I \cdot e^{(g_{_A} - d_{tvm}) \cdot t} 
\end{align}

The correct fit value for $d_{rar}$ (which cannot be known in advance) equates the expected stream to the actual stream. Call this $d_{rarF}$. To calculate it, we only care about the growth terms in eq. \ref{eq:Etvm_Atvm2}. Thus:
\begin{align}
g_{_E}-d_{tvm} - d_{rarF} &= g_{_A} - d_{tvm} \qquad 
\text{where: } 0 < g_A < g_E\\
g_{_E}-d_{rarF} &= g_{_A} \\
d_{rarF} &= g_{_E} - g_{_A}  > 0   
\end{align}

Therefore, because we are risk adjusting for $g_{_A}$ being lower than $g_{_E}$, $d_{rarF}$ is positive, as expected, and as accords with standard practice.

Now let us  consider a calamity, with an expected \emph{negative} stream of damages $D_E$ and an actual negative stream of damages $D_A$. The loss streams ($L)$ start at the same initial value $L<0$. These damages grow over time at the rates $g_{_E}$ and $g_{_A}$. And as before, a time value of money discount $d_{tvm}$ is applied to both loss streams, since losses at a future date are less concerning than losses in the present.
\begin{align}
D_E(t)&=L \cdot e^{(g_{_E}) \cdot t}   &  D_A(t)&=L \cdot e^{(g_{_A}) \cdot t} \\
D_{E_{tvm}}(t)&=L \cdot e^{(g_{_E} - d_{tvm}) \cdot t}   &  D_{A_{tvm}}(t)&=L \cdot e^{(g_{_A} - d_{tvm}) \cdot t} 
\end{align}
In the case of future losses, the risk is that these will be \emph{larger} than expected. We therefore seek to find a value of $d_{rar}$ to account for true losses of a larger magnitude than expected losses. As before, $d_{rar}$ is applied only to \emph{expected} losses, since $D_{A_{tvm}}(t)$ is the benchmark we seek to match:
\begin{align} \label{eq:Zero_gt_ActualLoss_gt_Expected}
\begin{split}
&D_{E_{rar}}(t) =L \cdot e^{(g_{_E} - d_{tvm} - d_{rar}) \cdot t} \qquad   D_{A_{rar}}(t) =  D_{A_{tvm}}(t)=L \cdot e^{(g_{_A} - d_{tvm}) \cdot t} \\ 
&\text{Where: } 0>D_{E}(t) > D_{A}(t) 
\end{split}
\end{align}

As in the positive income stream case, we are only interested in the growth terms of eq. \ref{eq:Zero_gt_ActualLoss_gt_Expected}. We are again interested in finding the fit (or actual) value for $d_{rar}$, which  is $d_{rarF}$.
\begin{align}
g_{_E}-d_{tvm} - d_{rarF} &= g_{_A} - d_{tvm} & \text{Where: } g_A < g_E < 0 \\
g_{_E}-d_{rarF} &= g_{_A} \\
d_{rarF} &= g_{_E} - g_{_A} < 0 
\end{align} 

Therefore, because the growth rate of actual damages, $g_{_A}$, exceeds those of expected damages $g_{_E}$, $d_{rarF}$ will be a \emph{negative} quantity. Since this term is subtracted from the exponent, the effect is to make the total discount smaller, not larger, when risk is taken into account. 

\section{Conclusion}
Stern is correct \cite{Stern2007EconomicsofClimateChangeSternReview} in his contention that after taking risk into account, a lower discount rate must be used. 


\bibliographystyle{plain} 
\bibliography{references} 

\end{document}